\documentclass{article}

\usepackage{amsmath}

\begin{document}
\title{A note on the geodesic deviation equation for null geodesics in the Schwarzschild black-hole}
\author{Juan J. Morales-Ruiz\thanks{Department of Applied Mathematics, Universidad Polit\'ecnica de Madrid, e-mail: \texttt{juan.morales-ruiz@upm.es}} \and \'Alvaro P. Raposo\thanks{Department of Applied Mathematics, Universidad Polit\'ecnica de Madrid, e-mail: \texttt{alvaro.p.raposo@upm.es}}}
\date{August 12, 2023}
\maketitle

\begin{abstract}
We use the Hamiltonian formulation of the geodesic equation in the Schwarzschild space-time so as to get the variational equation as the counterpart of the Jacobi equation in this approach.  In this context we are able to apply the Morales-Ramis theorem to link the integrability of the geodesic equation to the integrability, in the sense of differential Galois theory, of the variational equation.  This link is strong enough to hold even on geodesics for which the usual conserved quantities fail to be independent, as is the case of circular geodesics.  We show explicitly the particular cases of some null geodesics and their variational equations.
\end{abstract}

\section{Introduction}
The geodesic equation in a space-time $(M,g)$ is given by (see e.g.~\cite{Wal84})
\begin{equation}\label{eq:geodesic}
  \ddot x^\alpha + \Gamma_{\beta\gamma}^\alpha \dot x^\beta \dot x^\beta = 0,
\end{equation}
for $x^{\alpha}(\tau)$, a geodesic curve parametrized by an affine parameter $\tau$, where the four coordinates $(x^0, x^1, x^2, x^3)$ live in the space-time, that is, the four dimensional manifold $M$ provided with a semi-Riemannian metric $g$, from which the affine connection with coefficients $\Gamma^\alpha_{\beta\gamma}$ is constructed.  The dot denotes derivative with respect to the parameter of the curve, $\dot x^\alpha = \frac{{\rm d}x^\alpha}{{\rm d}\tau}$. In relativity theory, the geodesic equations model the motion of a test particle (i.e., one which does not alter the surrounding geometry) moving freely in a fixed space-time given by a solution of the Einstein field equations.

If the geodesic is timelike an affine reparametrization allows to interpret $\tau$ as proper time, in which case the normalization
\begin{equation}\label{eq:timelike}
  g_{\alpha\beta}\dot x^\alpha \dot x^\beta = -1
\end{equation}
holds. This is the case of material particles (those with non null mass), while massless test particles (like photons) follow null geodesics with the normalization
\begin{equation}\label{eq:null}
  g_{\alpha\beta}\dot x^\alpha \dot x^\beta = 0
\end{equation}
in which case the parameter cannot be interpreted as a proper time.

The equation of geodesic deviation, also known as Jacobi equation, is the linear differential equation of a vector $\xi^\alpha$ which connects a given geodesic with a nearby one, truncated to first order:
\begin{equation}\label{eq:Jacobi}
  \ddot \xi^\alpha + \Gamma^\alpha_{\beta\gamma}\dot x^\beta \dot\xi^\gamma + R^\alpha_{\beta\gamma\delta} \dot x^\beta \dot x^\delta \xi^\gamma = 0,
\end{equation}
where $R^\alpha_{\beta\gamma\delta}$ is the Riemann tensor evaluated on the given geodesic, as well as the velocity $\dot x^\alpha$ and the connection $\Gamma^\alpha_{\beta\gamma}$.  Notice the first two terms are usually written in compact form as $\frac{{\rm D}^2\xi^\alpha}{{\rm D} \tau^2}$, that is the covariant derivative along the curve.  This equation is interpreted in the context of relativity as modelling the movement of a set of structureless test particles freely falling in the surrounding gravitational field. The deviation vector $\xi^\alpha$ describes the relative position of such a particle respect to a reference one and, thus, describe if they approach each other or get apart due to the tidal gravitational forces described by the Riemann tensor.

The study of the integrability of both, the geodesic equation~\eqref{eq:geodesic} and the Jacobi equation~\eqref{eq:Jacobi}, have been classically based on Killing vectors and, more generally, tensors.  A Killing vector $k^\alpha$ is the generator of a one-parameter group of isometries of the metric and, thus, satisfies the relation
\begin{equation}\label{eq:Killing-vector}
  \nabla_\beta k_\alpha + \nabla_\alpha k_\beta = 0,
\end{equation}
where $\nabla$ is the covariant derivation associated with the metric $g$. Each Killing vector produces a first integral of the geodesic (i.e. a conserved quantity along the geodesic curve) as
\begin{equation}\label{eq:K}
  K=k_\alpha \dot x^\alpha.
\end{equation}
A Killing tensor is defined by a generalization of equation~\eqref{eq:Killing-vector} as a totally symmetric tensor $k^{\alpha_1 \dots \alpha_r}=k^{(\alpha_1 \dots \alpha_r)}$ which satisfies
\begin{equation}\label{eq:Killing-tensor}
  \nabla_{(\beta}k_{\alpha_1 \dots \alpha_r)}=0,
\end{equation}
where the brackets stand for symmetrization, from which the quantity $K=k_{\alpha_1 \dots \alpha_r}\dot x^{\alpha_1}\dots \dot x^{\alpha_r}$ is conserved along the geodesic.

The geodesic equations constitute a dynamical system of 4 degrees of freedom and its integrability is usually understood as the exhibition of 4 conserved quantities emerging from 4 independent Killing tensors.

The integrability of the Jacobi equation has also been studied from the same point of view.  A recent work by Cariglia et al.~\cite{Car18} proves that the complete integrability of the geodesic equation leads to the complete integrability of the Jacobi equation.  The result is based on the same Killing tensors which also produce conserved quantities for the solutions of the Jacobi equation.  H. Fuchs~\cite{Fuc77} established that, if $k^\alpha$ is a Killing vector, then $k_\alpha \frac{{\rm D}\xi^\alpha}{{\rm D}\tau} - \xi_\alpha \frac{{\rm D}k^\alpha}{{\rm D}\tau}$ is a first integral of the Jacobi equation, and uses it succesfully to completely integrate the Jacobi equation in the Schwarzschild spacetime~\cite{Fuc83}.  Another approach is given by Ba\.za\'nski and Jaranowski~\cite{Baz89} in terms of a complete integal $U$ of the Hamilton-Jacobi equation from which the needed conserved quantities are built, and they also employ their method to completely integrate the Jacobi equation in the Schwarzschild spacetime.  However, all these works limit themselves to the Jacobi equation around timelike geodesics.  Moreover, the method breaks down in some particular scenarios, such as circular geodesics in the Schwarzschild spacetime, because the killing vectors are not linearly independent (see section~\ref{subsec:circular}); for instance, H. Fuchs devotes a different paper~\cite{Fuc90} to the integrability of the Jacobi equation around timelike circular geodesics.

In this paper we propose another point of view of the integrability issue of the Jacobi equation starting from the equivalent formulation of the geodesics as a Hamiltonian system.  In section~\ref{sec:Hamiltonian-formulation} we remind this formulation of the geodesic equation and the associated variational equation.  For the Hamiltonian system we use Liouville's integrability definition of $n$ independent conserved quantities in involution, which are the same as the conserved quantities obtained from the Killing vectors.  However, for the variational equation we use the integrability criterion given by differential Galois theory so we are allowed to make use of the Morales-Ramis theorem to relate both integrabilities. This approach has been used in the past to study the integrability of geodesic flows in a variety of manifolds, see for instance~\cite{Com15} and references therein. In section~\ref{sec:Schwarzschild} we apply this relation to the geodesics in Schwarzchild space-time, with explicit computation in two families of null geodesics: circular geodesic and radial geodesic.  The former is an instance in which the usual approach by means of Killing vectors fails, but is straighforward via Morales-Ramis theorem.  Finally, section~\ref{sec:conclusions} summarizes our conclusions.

Our aim to this approach is a program of reviewing integrable classical Hamiltonian systems.  The associated variational equation of a Hamiltonian system is the basis for a semiclassical approximation to its quantization (that is an actual instance of the statement by Bryce DeWitt ``quantum theory is basically a theory of small disturbances'').   By means of Morales-Ramis theorem, these systems lead to integrable variational equations~\cite{Mor20}.  Therefore, integrable classical Hamiltonian systems produce semiclassical quantum approximations which can be explicitly computed in closed form.

\section{Hamiltonian formulation and integrability}\label{sec:Hamiltonian-formulation}
  It is well known that the geodesics can be formulated as a Hamiltonian system.  With the same variables $x^\alpha(\tau)$, their conjugated momenta are $p_\alpha = \frac{1}{2} g_{\alpha\beta}\dot x^\beta$ and the Hamiltonian function then reads
\begin{equation}\label{eq:Hamiltonian}
  H(x^\alpha, p_\alpha) = \frac{1}{2} g^{\alpha\beta}p_\alpha p_\beta,
\end{equation}
where the dependence on the variables $x^\alpha$ is given through the metric $g$.
The Hamilton equations
\begin{subequations}\label{eq:Hamilton-equations}
  \begin{align}
  \dot x^\alpha &= \frac{\partial H}{\partial p_\alpha} = g^{\alpha\beta}p_\beta, \label{eq:Hamilton-equations-x}\\
  \dot p_\alpha &= -\frac{\partial H}{\partial x^\alpha} = -\frac{1}{2} g^{\beta\gamma}_{\hspace{5pt},\alpha} p_\beta p_\gamma, \label{eq:Hamilton-equation-p}
  \end{align}
\end{subequations}
are equivalent to the geodesic equations~\eqref{eq:geodesic}.  As it is well known, integrability in this setting is Liouville integrability, i.e. four functions of $(x^\alpha, p_\alpha)$ algebraically independent and in involution (that is, each pair of them Poisson commute).  The Hamiltonian function is one of these functions since it does not depend explicitly on the parameter $\tau$.  In fact, its value, due to the restrictions~\eqref{eq:timelike} and \eqref{eq:null} is $H=-1/2$ for  timelike geodesics and $H=0$ for null geodesics.

The conserved quantity $K$ of equation~\eqref{eq:K} derived from a Killing vector traslates as a conserved quantity in this setting as can be verified directly; the Poisson bracket $\{K,H\}=0$ thanks to the Killing equation~\eqref{eq:Killing-vector}.  It is also well known that two linearly independent Killing vectors which commute give rise to two algebraically independent conserved quantities in involution. Therefore, integrability in the Liouville sense within the Hamiltonian formulation is equivalent to integrability of the geodesic equation via Killing vectors.
  
For dynamical systems such as Hamiltonian ones, Poincar\'e introduced the variational equation, which is the linear equation on the variations $(\xi^\alpha, \eta_\alpha)$ which gives a solution, to first order, to the Hamilton equations in the form $(x^\alpha + \xi^\alpha, p_\alpha+\eta_\alpha)$, where $(x^\alpha, p_\alpha)$ is a known geodesic.  The variational equations read
\begin{subequations}\label{eq:variational-equations}
  \begin{align}
    \dot\xi^\alpha &= \frac{\partial^2 H}{\partial p_\alpha \partial x^\beta}\, \xi^\beta + \frac{\partial^2 H}{\partial p_\alpha \partial p_\beta}\, \eta_\beta =
    g^{\alpha\gamma}_{\hspace{5pt},\beta} p_\gamma \,\xi^\beta + g^{\alpha\beta} \,\eta_\beta, \label{eq:variational-equations-xi}\\
    \dot\eta_\alpha &= -\frac{\partial^2 H}{\partial x^\alpha \partial x^\beta}\, \xi^\beta - \frac{\partial^2 H}{\partial x^\alpha \partial p_\beta}\, \eta_\beta=
    -\frac{1}{2}g^{\gamma\delta}_{\hspace{5pt},\alpha\beta} p_\gamma p_\delta \,\xi^\beta - g^{\beta\gamma}_{\hspace{5pt},\alpha} p_\gamma \,\eta_\beta, \label{eq:variational-equations-eta}
  \end{align}
\end{subequations}
where the second order derivatives of the Hamiltonian function are computed on the geodesic.

Equations~\eqref{eq:variational-equations} are equivalent to the Jacobi equation~\eqref{eq:Jacobi}.  To see this equivalence, start by deriving with respect to $\tau$ in equation~\eqref{eq:variational-equations-xi} and take into account that both $g^{\alpha\beta}$ and $p_\alpha$ depend on $\tau$.  Thus, a factor $\dot p_\alpha$ is substituted by its expression as given by equation~\eqref{eq:Hamilton-equation-p} and with $p_\alpha$ substituted in terms of $x^\alpha$ and $\dot x^\alpha$ by means of equation~\eqref{eq:Hamilton-equations-x}.  The derivative of $g^{\alpha\beta}$ is given by $\frac{\rm d}{{\rm d}\tau} g^{\alpha\beta} = g^{\alpha\beta}_{\hspace{1em},\gamma}\dot x^\gamma$.  In addition, the variable $\eta_\alpha$ is to be substituted by an expression with only $\xi^\alpha$ and $\dot \xi^\alpha$ by means of equation~\eqref{eq:variational-equations-xi} as well as the factor $\dot \eta_\alpha$ is similarly substituted by means of equation~\eqref{eq:variational-equations-eta}.  The resulting intermediate equation is
\begin{multline}\label{eq:intermediate}
  \ddot \xi^\alpha = 
  \left ( g^{\alpha\beta}_{\hspace{5pt},\gamma\mu}g_{\beta\nu} - \frac{1}{2} g^{\alpha\beta}_{\hspace{5pt},\gamma} g^{\sigma\delta}_{\hspace{5pt},\beta} g_{\mu\sigma}g_{\delta\nu} + g^{\alpha\beta}_{\hspace{5pt},\mu}g_{\beta\nu,\gamma} - \frac{1}{2} g^{\alpha\beta}g^{\epsilon\delta}_{\hspace{5pt},\beta\gamma} g_{\epsilon\mu}g_{\delta\nu}    \right ) \dot x^\mu \dot x^\nu \xi^\gamma \\
  + \left ( g^{\alpha\beta}_{\hspace{5pt},\gamma}g_{\beta\delta} + g^{\alpha\beta}_{\hspace{5pt},\delta}g_{\beta\gamma} - g^{\alpha\beta}g^{\epsilon\mu}_{\hspace{5pt},\beta} g_{\delta\mu}g_{\gamma\epsilon}    \right ) \dot x^\delta \dot \xi^\gamma.
\end{multline}
The next step is to recognize the Riemann tensor and the Christoffel symbol in the first and second expressions in brackets, respectively, by writting both, $R^\alpha_{\beta\gamma\delta}$ and $\Gamma^\alpha_{\beta\gamma}$ in terms of the metric tensor $g_{\alpha\beta}$ and its derivatives and, thus, arrive at equation~\eqref{eq:Jacobi}.

The other way, from the Jacobi equation to the variational equation, also holds, so the variables $\xi^\alpha$ are the same in both equations and one can study it through any of them.

Integrability of the Jacobi equation or, equivalently, the variational equation, is where we are introducing a different point of view.

\subsection{Integrability of variational equation as per differential Galois theory}
So far we have just stated the problem with an alternative formulation, but nothing different has emerged.  Now we wish to use a different point of view to the integrability of the linear equation on the deviation vector.  We wish to use the integrability criterion defined within differential Galois theory, which is more precise and insightful than Liouville's.  While the latter states that a system is integrable if there are enough conserved quantities from which to ``integrate'' the variables, differential Galois criterion is more detailed. The differential Galois theory of linear ordinary differential equations (see, e.g.~\cite{Van03, Mor13}) focus on the coefficients of the linear differential equation, which are defined on a differential field, and on the solutions to the equation, which may be defined on another differential field, extension of the former.  The notion of integrability relies in the structure of the extension differential field with respect to the differential field of coefficients.  If the extension can be generated from the original by incorporating either (1) algebraic functions (those which satisfy an algebraic equation on the original differential field), (2) primitive functions (i.e. their derivatives are in the original differential field) or (3) exponentials of primitives in the original differential field, then the equation is said to be integrable.

In the same vein as classical Galois theory of polynomials, an extension of differential fields defines a group of differential automorphisms, the differential Galois group.  One of the main results of this theory is the theorem which states that a linear differential equation is integrable if and only if the identity component of its Galois group is solvable.

With respect to the problem posed by the geodesic equation and the linear equation of small disturbances around a given geodesic, Galois theory has produced a powerful result, the so called Morales-Ramis theorem~\cite{MR01} (see also~\cite{Mor13}).  Roughly speaking this theorem states that the integrability of a Hamiltonian system, in the sense of Liouville, implies the integrability of the associated variational equation (in fact, it implies the identity component of its Galois group is commutative).

Traditionally this theorem has been used as a non-integrability result.  If the variational equation of a Hamiltonian system is non-integrable from the Galoisian point of view, then the system itself is not integrable in Liouville's sense.  However, in our case, the theorem can be applied in the direct way.  The advantage is that this theorem includes the particular cases where the integrability of the geodesic equation can be compromised, for instance if the conserved quantities become algebraically dependent on the geodesic curve, as is the case of the circular geodesics in Schwarzschild space-time.

\section{Schwarzschild geodesics}\label{sec:Schwarzschild}
We now focus on the Schwarzschild space-time and its geodesic curves, firstly, and their variational equation, secondly.  We employ the usual coordinates $x^\mu = (t,r,\theta, \varphi)$, which are interpreted as coordinate time $t$ and spherical coordinates $r, \theta, \varphi$.  The metric tensor reads
\begin{equation}\label{eq:metric}
  g_{\alpha\beta} = \begin{pmatrix}
    -\left( 1-\frac{2M}{r} \right) & 0 & 0 & 0 \\
    0 & \frac{1}{1-\frac{2M}{r}} & 0 & 0 \\
    0 & 0 & r^2 & 0 \\
    0 & 0 & 0 & r^2 (\sin \theta)^2 
  \end{pmatrix},
\end{equation}
where $M$ is interpreted as the mass of the object which creates the gravitational field (a star, black hole, etc.), in a system of units where $G=c=1$, located at the origin of spherical coordinates.  These coordinates are well known for breaking at $r=2M$, the Schwarzschild radius, which is just an apparent singularity, but has the physical interpretation of the event horizon of a static and sphericallly symmetric black hole.

The Hamiltonian function of equation~\eqref{eq:Hamiltonian} under this metric tensor reads
\begin{equation}\label{eq:Schwarzschild-Hamiltonian}
  H(x^\alpha, p_\alpha) = \frac{-1}{2(1-\frac{2M}{r})} p_t^2 + \frac{1}{2} \left (1-\frac{2M}{r} \right ) p_r^2 + \frac{1}{2 r^2} p_\theta^2 + \frac{1}{2 r^2 (\sin \theta)^2} p_\varphi^2,
\end{equation}
and the subsequent Hamilton equations are
\begin{subequations}\label{eq:Schwarzschild-geodesics}
\begin{align}
    \dot t &= \frac{-1}{1-\frac{2M}{r}} p_t, &  \dot p_t &= 0, \label{eq:Schwarzschild-geodesics-t}\\
    \dot r &= \left( 1-\frac{2M}{r} \right ) p_r, &     \dot p_r &= \frac{-M}{(r-2M)^2} p_t^2 + \frac{M}{r^2} p_r^2 + \frac{1}{r^3} p_\theta^2 + \frac{1}{r^3(\sin \theta)^2} p_\varphi^2, \label{eq:Schwarzschild-geodesics-r}\\
    \dot \theta &= \frac{1}{r^2} p_\theta, &     \dot p_\theta &= \frac{\cos \theta}{r^2(\sin \theta)^3} p_\varphi^2, \label{eq:Schwarzschild-geodesics-theta}\\
    \dot \varphi &= \frac{1}{r^2(\sin \theta)^2} p_\varphi, & \dot p_\varphi &= 0. \label{eq:Schwarzschild-geodesics-phi}
\end{align}
\end{subequations}
Since $t$ and $\varphi$ are cyclic variables we have readily the following constants of motion: $E=-p_t = \left ( 1-\frac{2M}{r} \right ) \dot t$, the energy of the particle (with a choice of sign such that $E>0$ means $\dot t > 0$ in the region $r>2M$, that is, the coordinate time grows along with the parameter $\tau$; $L=p_\varphi=r^2(\sin\theta)^2\dot\varphi$, the angular momentum of the particle. Also the Hamiltonian function~\eqref{eq:Schwarzschild-Hamiltonian} is conserved as stated previously.

In addition, the polar axis can be chosen arbitrarily, for instance such that the initial conditions of the geodesic are $\theta_0=\pi/2$ and $\dot\theta_0=0$.  Then, the equation for the conjugate momentum gives $p_\theta = 0$, constant and, thus, $\theta = \pi/2$, constant as well. We have at our disposal five constants of motion, although only four of them are in involution, say $(H, E=-p_t, L=p_\varphi, \theta)$.  And they are algebraically independent: $\theta$, $p_t$ and $p_\varphi$ are independent variables in the Hamiltonian formulation, and $H$ is also independent from the other three since it has a term with $p_r$, which is a different variable (this will be relevant when we come to circular geodesics, in which $p_r=0$).  Therefore, as it is well known, the geodesic equation in the Schwarzschild spacetime is an integrable system.

Now, the Morales-Ramis theorem informs us of the integrability of the variational equations~\eqref{eq:variational-equations} in the sense of differential Galois theory, for the Schwarzschild geodesics.  The variational equations for the eight variations $\xi^t$, $\xi^r$, $\xi^\theta$, $\xi^\varphi$, $\eta_t$, $\eta_r$, $\eta_\theta$, $\eta_\varphi$ are:
\begin{subequations}\label{eq:Schwarzschild-variational-equation}
\begin{align}
    \dot \xi^t &= H_{p_tr}\xi^r + H_{p_tp_t} \eta_t, &     \dot \eta_t &= 0, \label{eq:Schwarzschild-variational-equation-t}\\
    \dot \xi^r &= H_{p_rr}\xi^r + H_{p_rp_r} \eta_r, &     \dot \eta_r &= -H_{rr} \xi^r - H_{rp_t} \eta_t - H_{rp_r} \eta_r - H_{rp_\varphi} \eta_\varphi, \label{eq:Schwarzschild-variational-equation-r}\\
    \dot \xi^\theta &= H_{p_\theta p_\theta}\eta_\theta, &     \dot \eta_\theta &= -H_{\theta\theta} \xi^\theta, \label{eq:Schwarzschild-variational-equation-theta}\\
    \dot \xi^\varphi &= H_{p_\varphi r}\xi^r + H_{p_\varphi p_\varphi} \eta_\varphi, &     \dot \eta_\varphi &= 0, \label{eq:Schwarzschild-variational-equation-phi}
\end{align}
\end{subequations}
where the elements of the Hessian matrix of the Hamiltonian function written in the equations are the only ones which are nonzero, and they are evaluated on the particular geodesic whose variations are under study.

Notice the similarities between the variational equations~\eqref{eq:Schwarzschild-variational-equation} and the geodesic equations~\eqref{eq:Schwarzschild-geodesics} which makes evident in this case the relation between the integrability of one system and the other.

The complete solution of both sets of equations, the geodesic equations and the variational equations (in the form of Jacobi equation) are well known.  Instead we are interested in some particular cases of geodesics which have not been covered by previous studies.  We now focus on null geodesics and, in particular, circular and radial null geodesics.

\subsection{Variational equation around some Schwarzschild null geodesics}
\subsubsection{Circular null geodesics}\label{subsec:circular}
There are solutions for the geodesics equations~\eqref{eq:Schwarzschild-geodesics-t} to \eqref{eq:Schwarzschild-geodesics-phi} with $H=0$ (null geodesic) and $\dot r=0$ (circular, but $r\not= 2M$, where the coordinates fail).  Upon substitution on the geodesic equations it is straightforward to get the well known circular null geodesics (which define the photons sphere) described by
\begin{subequations}\label{eq:circular-null-geodesics}
\begin{align}\label{eq:circular-null-geodesics-t}
    t(\tau)&=3E \tau + t_0,  & p_t(\tau)&=-E, \\
    r(\tau)&=3M,  & p_r(\tau)&=0, \\
    \theta(\tau)&=\frac{\pi}{2}, & p_\theta(\tau)&=0, \\
    \varphi(\tau) &= \frac{E}{\sqrt{3}M} \tau + \varphi_0, & p_\varphi(\tau) &= L,\label{eq:circular-null-geodesics-phi}
\end{align}
\end{subequations}
where the parameter $E > 0$ characterizes each solution within this family because $L$ and $E$ are related by $L^2=27M^2E^2$.  This fact means that our four conserved quantities are not independent, a fact that comes from the restriction $p_r=0$ which leaves the Hamiltonian function of equation~\eqref{eq:Schwarzschild-Hamiltonian} depending only on $p_t$ and $p_\varphi$, so $H$, $p_t$ and $p_\varphi$ are no longer algebraically independent.  In terms of killing vectors, the killing vectors $\frac{\partial}{\partial t}$ and $\frac{\partial}{\partial \varphi}$ are not linearly independent.  However, this relation holds only on the geodesic curve, but not around it, which is precisely where the deviation field $\xi^\alpha$ operates.  This possibility is accounted for in the Morales-Ramis theorem, which ask for conserved quantities algebraically independent and in involution except, possibly, on the integral curve.  So, with this approach, we do not have to split the study of this particular case from the general one.  Of course the geodesic equation has been integrated despite this reduction of constants because we have introduced another one, namely $r(\tau)$, but this is not the case in the variational equation.  Nevertheless, as we have said, the Morales-Ramis theorem asserts that it is also integrable in this case.  Let us see how it works.

Taking the solution~\eqref{eq:circular-null-geodesics} as the basis for the variations, the variational equations~\eqref{eq:Schwarzschild-variational-equation-t} to \eqref{eq:Schwarzschild-variational-equation-phi} take the form
\begin{subequations}\label{eq:circular-variational-equations}
  \begin{align}
    \dot\xi^t &= -\frac{2E}{M} \xi^r - r \eta_t \\
    \dot\xi^r &= \frac{1}{3} \eta_r \\
    \dot\xi^\theta &= \frac{1}{9M^2} \eta_\theta \\
    \dot\xi^\varphi &= -\frac{2E}{\sqrt{27}M^2} \xi^r + \frac{1}{9M^2} \eta_\varphi \\
    \dot\eta_t &= 0 \\
    \dot\eta_r &= \frac{E^2}{M^2} \xi^r + \frac{2E}{M} \eta_t + \frac{2E}{\sqrt{27}M^2} \eta_\varphi \\
    \dot\eta_\theta &= -3E^2 \xi^\theta \\
    \dot\eta_\varphi &= 0.
  \end{align}
\end{subequations}
This system is straighforwardly solvable by noticing that the coefficients are constant. The complete solution of the variational equation in closed form can be written as
\begin{subequations}
\begin{align}\label{eq:circular-variation}
    \xi^t(\tau) &= \left (\frac{6M}{E} - 2\sqrt{3}C_1 - 2\sqrt{3} C_2 \tau \right ) e^{\frac{E}{\sqrt{3}M} \tau} + \left(9D_1 + \frac{4}{\sqrt{3}M} D_2 \right ) + B_1,\\
    \xi^r(\tau) &= (C_1 + C_2 \tau) e^{\frac{E}{\sqrt{3}M} \tau} - \frac{6M}{E} D_1 - \frac{6}{\sqrt{27}E} D_2,\\
    \xi^\theta(\tau) &= A_1 \sin \left( \frac{E\tau}{\sqrt{3}M}\right) + A_2 \cos \left( \frac{E\tau}{\sqrt{3}M}\right), \\
    \begin{split}
    \xi^\varphi(\tau) &= \left (\frac{2}{\sqrt{3}E} - \frac{2}{3M}C_1 - \frac{2}{3M} C_2 \tau \right ) e^{\frac{E}{\sqrt{3}M} \tau} \\ &+ \left(\frac{4}{\sqrt{3}M}D_1 + \frac{12\sqrt{3}+1}{9M^2} D_2 \right )\tau + B_2,
    \end{split}    \\
    \eta_t(\tau) &= D_1, \\
    \eta_r(\tau) &= \left (\frac{\sqrt{3}E}{M} C_1 + 3C_2 -\frac{\sqrt{3}E}{M} C_2 \tau \right) e^{\frac{E}{\sqrt{3}M} \tau}, \\
    \eta_\theta(\tau) &= 3\sqrt{3} EM \left ( A_1 \cos \left( \frac{E\tau}{\sqrt{3}M}\right) - A_2 \sin \left( \frac{E\tau}{\sqrt{3}M}\right) \right), \\
    \eta_\varphi(\tau) &= D_2,
\end{align}
\end{subequations}
where $A_1$, $A_2$, $B_1$, $B_2$, $C_1$, $C_2$, $D_1$, $D_2$ are the eight arbitrary integration constants to be determined by the initial conditions.

We now compare the form of the explicit solution with the coefficients of the variational equations~\eqref{eq:circular-variational-equations}.  We see that the solution is made of polynomials and exponential functions $e^{\frac{E}{\sqrt{3}M} \tau}$, so it is an allowed extension within the integrability criterion of differential Galois theory.  The Galois group of this differential field extension is isomorphic with the multiplicative group of complex numbers, that is, a commutative group as predicted by the Morales-Ramis theorem since the group is connected.

\subsubsection{Radial null geodesics}
There are also solutions for the geodesic equations~\eqref{eq:Schwarzschild-geodesics} with $H=0$ (null) and $\varphi$ constant, i.e. $L=0$, (radial).  A straightforward computation after substitution on the geodesics equations gives the well known radial null geodesics as
\begin{align}\label{eq:radial-null-geodesics}
    t(\tau)&=E \tau + 2M \log \left( 1 + \frac{E\tau}{r_0-2M} \right) + t_0,  & p_t(\tau)&=-E, \\
    r(\tau)&=\pm E\tau + r_0,  & p_r(\tau)&=\pm \frac{E(E\tau + r_0)}{E\tau +r_0 - 2M}, \\
    \theta(\tau)&=\frac{\pi}{2}, & p_\theta(\tau)&=0, \\
    \varphi(\tau) &= \varphi_0, & p_\varphi(\tau) &= L = 0.
\end{align}
The plus sign refers to an outgoing geodesic ($\dot r > 0$) and the minus sign to an ingoing geodesic ($\dot r < 0$).

The variations $\xi^\alpha$ and $\eta_\alpha$ around the solution~\eqref{eq:radial-null-geodesics} satisfy the following equations:
\begin{subequations}\label{eq:radial-variational}
  \begin{align}
    \dot \xi^t &= \frac{-2ME}{(r-2M)^2} \xi^r - \frac{r}{r-2M} \eta_t, \\
    \dot \xi^r &= \frac{2ME}{r(r-2M)} \xi^r + \frac{r-2M}{r} \eta_r, \\
    \dot \xi^\theta &= \frac{1}{r^2} \eta_\theta, \\
    \dot \xi^\varphi &= \frac{1}{r^2} \eta_\varphi, \\
    \dot \eta_t &= 0, \\
    \dot \eta_r &= \frac{4ME^2(r-M)}{r(r-2M)^3} \xi^r - \frac{2ME}{r(r-2M)} \eta_r + \frac{2ME}{(r-2M)^2} \eta_t, \\
    \dot \eta_\theta &= 0, \\
    \dot \eta_\varphi &= 0.
\end{align}
\end{subequations}
where, in the coefficients, $r$ must be substituted by its expression in eq.~\eqref{eq:radial-null-geodesics}, therefore the coefficients are not constant. Nevertheless this system can be solved by splitting the decoupled equations ($\xi^\theta$ and $\eta_\theta$, $\xi^\varphi$ and $\eta_\varphi$ and, finally, $\xi^r$ and $\eta_r$).
The complete solution of the variational equation in closed form is
\begin{subequations}\label{eq:radial-variational-solution}
  \begin{align}
    \begin{split}
      \xi^t(\tau) &= -D_1 \tau + \left( 2MC_2 - (C_1+D_1) \frac{2M(r_0-2M)}{E}\right) \frac{1}{E\tau +r_0-2M} \\
        & -\frac{2M}{E} C_1 \log\left( \frac{E\tau + r_0 - 2M}{2M}\right) - \frac{C_1}{E^2} \log\left( \frac{E\tau + r_0 - 2M}{E\tau + r_0}\right) \\
        & +\frac{2M}{E^2} C_1 \frac{\log\left( \frac{E\tau + r_0 - 2M}{2M}\right)}{E\tau + r_0 -2M} + B_1,
    \end{split} \\
    \xi^r(\tau) &= C_1 \left( \tau - \frac{2M}{E} \log \left( \frac{E\tau +r_0}{2M}\right)\right) - D_1 \tau + C_2, \\
    \xi^\theta(\tau) &= -\frac{A_2}{E(E\tau + r_0)} + A_1,\\
     \xi^\varphi(\tau) &= -\frac{D_2}{E(E\tau + r_0)} + B_2,\\
  \eta_t(\tau) &= D_1, \\
\begin{split}
  \eta_r(\tau) &=  (C_1 - 2MEC_2) -D_1 \frac{E \tau +r_0}{E\tau+r_0-2M} \\
   & - 2ME (C_1-D_1) \frac{\tau}{(E\tau+r_0-2M)^2} + 4M^2 \frac{\log \left( \frac{E\tau + r_0}{2M} \right)}{(E\tau+r_0-2M)^2},
\end{split} \\
    \eta_\theta(\tau) &= A_2, \\
    \eta_\varphi(\tau) & =D_2,
\end{align}
\end{subequations}
with $A_1$, $A_2$, $B_1$, $B_2$, $C_1$, $C_2$, $D_1$ and $D_2$ the eight arbitrary integration constants to be determined by the initial conditions.

We now notice that the coefficients of the equations~\eqref{eq:radial-variational} are rational functions in the parameter $\tau$.  The solution~\eqref{eq:radial-variational-solution} is made of rational functions on $\tau$ and logarithmic functions on $\tau$, which is a class of the so called primitive functions, for its derivative belongs to the original differential field of rational functions in $\tau$.  Therefore the extension is again an integrable one.  The Galois group associated with this extension is isomorphic with a direct product with copies of the additive group of complex numbers, i.e. a commutative group as predicted by the Morales-Ramis theorem since, once again, it is a connected group.

\section{Conclusions}\label{sec:conclusions}
We have shown that the relation between integrability of the geodesic equation and that of the Jacobi equation in a spacetime can be understood with more generality by means of a Hamiltonian formulation of the geodesic equations and, thus, the variational equation instead of the Jacobi equation (of which the former is equivalent).

With this formulation the Morales-Ramis theorem can be applied, with the advantage that it holds also in the cases where the integrability of the Hamiltonian system does not hold because the conserved quantities fail to be independent.  As an instance, this is the case of the circular geodesic, which in the past had to be treated separately, but no more from this point of view.

The authors wish to acknowledge Jean Pierre Ramis for stimulating discussions and the research group Non-linear Mathematical Models of the Universidad Polit\'ecnica de Madrid.

\bibliographystyle{plain}
\bibliography{MoRa1_2023}

\end{document}